\documentclass[11pt]{article}

\usepackage{acl}

\usepackage{times}
\usepackage{latexsym}
\usepackage{amsmath}
\usepackage{fontspec}
\usepackage{polyglossia}
\usepackage{booktabs}
\usepackage{tabularx}
 \usepackage{multirow}

\setmainlanguage{english}
\setotherlanguage{arabic}

\newfontfamily\arabicfont[
  Script = Arabic,
  Path = fonts/,
  Extension = .ttf,
  UprightFont = *-Regular,
  BoldFont = *-Bold,
  ItalicFont = *-Italic,
  BoldItalicFont = *-BoldItalic
]{Amiri}

\usepackage{newtxtext}
\usepackage{rotating}

\usepackage{microtype}

\usepackage{inconsolata}
\usepackage{amsmath}

\usepackage{graphicx}

\title{What Makes a Good Fiqh Retriever?

Answer Retrieval for Arabic Islamic Jurisprudence}

\author{Somaya Eltanbouly, Heba Sbahi, Samer Rashwani \\ {\bf Abdessalam Bouchekif, Mutaz Al-Khatib, Shahd Gaben, Mohammed Ghaly} \\
  Hamad Bin Khalifa University, Qatar \\
\texttt{\{seltanbouly, hsbahi, srashwani, abouchekif, malkhatib, sgaben, mghaly\}@hbku.edu.qa}
}

\begin{document}
\maketitle
\begin{abstract}
Retrieval-Augmented Generation is used for Islamic question answering, but most systems are evaluated end-to-end, making retrieval failures difficult to isolate from generation failures. We study answer-bearing retrieval for Arabic fiqh, where a passage is relevant only if it states the ruling required by the question. We build a retrieval test collection for Arabic fiqh and use it to evaluate dense, lexical, hybrid, fine-tuned, and madhhab-aware retrieval strategies. The best retriever achieves 0.524 MRR@5, while fine-tuning improves performance to 0.553. Hybrid retrieval provides limited gains for strong models, whereas madhhab-aware filtering more than doubles MRR@5 on school-specific questions. We further present an error analysis showing that the main challenge is distinguishing answer-bearing passages from topically similar passages that do not contain the requested ruling. The best-performing model and fine-tuning dataset are publicly available at \url{https://huggingface.co/collections/somayaeltanbouly/badr-arabic-fiqh-retrieval}.
\end{abstract}

\section{Introduction}
Retrieval-Augmented Generation (RAG) is increasingly used in Islamic question answering because religious rulings must be grounded in trustworthy and verifiable sources \cite{fanar-sadiq}. This is particularly important in fiqh, where the same question may have different valid rulings depending on the madhhab \cite{Sacred-Synthetic}. If retrieval returns an incomplete, irrelevant, or school-inconsistent passage, the generated answer may be misleading even when the language model itself is fluent. Recent work has applied RAG to fiqh question answering \cite{Aftina, bhatia2026ragagenticragfaithful, alan2025improvingllmreliabilityrag} and Quran-related tasks \cite{Aqrag, Khalila2025}, demonstrating the potential of combining retrieval with Large Language Models (LLMs) for religious question answering.

Despite this progress, most Islamic RAG systems are evaluated end-to-end, making it difficult to distinguish retrieval failures from generation failures \cite{fanar-sadiq, alan2025improvingllmreliabilityrag, Aftina, rag-survey}. This limitation is particularly important for fiqh question answering, where a generation model can only produce a grounded answer if the retrieved context contains the ruling needed to answer the question.

Retrieval in fiqh also poses challenges that are not captured by standard retrieval evaluation. For question answering, a passage may discuss the correct topic without providing the ruling required by the question. Furthermore, for madhhab-specific questions, a retrieved passage must not only contain the ruling but also correspond to the correct jurisprudential school. We therefore evaluate retrieval using \emph{answer-bearing relevance}, where a passage is considered relevant only if it contains the ruling needed to answer the question.

Motivated by these challenges, we study the retrieval component of a fiqh RAG system. We build a retrieval test collection for Arabic fiqh and use it to evaluate dense, lexical, hybrid, fine-tuned, and madhhab-aware retrieval strategies. We investigate two research questions: \textbf{RQ1}: What retrieval configuration is most effective for answer-bearing fiqh retrieval? and \textbf{RQ2}: How does madhhab information affect fiqh retrieval?

Our contributions are fourfold. First, we introduce the first answer-bearing and madhhab-aware retrieval evaluation for Arabic fiqh, where relevance requires stating the requested ruling rather than merely matching the topic. Second, we construct a fiqh retrieval test collection covering both general and madhhab-specific questions. Third, we evaluate a diverse set of retrieval models and strategies, including multilingual, Arabic-focused, domain-adapted, fine-tuned, hybrid, and madhhab-aware retrievers. Finally, we provide a quantitative and qualitative analysis of retrieval failures in answer-bearing fiqh retrieval.

The remainder of this paper is organized as follows. Sections~\ref{bg} and~\ref{rl} provide background about Islamic  Jurisprudence and review related work, respectively. Section~\ref{pipeline} describes the retrieval pipeline, and Section~\ref{ES} presents the experimental setup. Section~\ref{results} reports the evaluation results, followed by the error analysis in Section~\ref{EA}. Finally, Section~\ref{conclusion} concludes the paper.

\section{Background} \label{bg}

Islamic jurisprudence (\textit{fiqh}) refers to the scholarly formulation of
practical legal rulings through the interpretation of Islamic sources. Classical fiqh literature is organized around legal questions (\textit{mas\={a}'il}) covering different areas such as worship, transactions, and family law.

A central concept in fiqh is the \textit{madhhab}, or jurisprudential school. The four major Sunni schools-\d{H}anaf\={i}, M\={a}lik\={i}, Sh\={a}fi'i, and \d{H}anbal\={i}-developed distinct methodological traditions, and the ruling for the same legal question may therefore differ across schools. Classical works are typically written within the framework of a particular madhhab, although they may also cite or discuss opinions from other schools. Consequently, identifying a passage as topically related to a question is not necessarily sufficient for retrieval, particularly when a query asks for the ruling of a specific madhhab or with specific conditions.

These characteristics make fiqh retrieval different from conventional semantic or topical retrieval. Passages concerning the same legal issue can have high lexical and semantic similarity while stating different rulings, conditions, or school-specific positions. Conversely, an answer-bearing passage may use terminology different from that of the user question. In this work, we therefore evaluate retrieval based on whether a passage explicitly provides the ruling requested by the query, rather than whether it merely discusses the same topic.

\section{Related Work} \label{rl}
Recently, many Islamic AI applications have been built using RAG pipelines to reduce hallucinations and improve the grounding of generated responses. For example, several participants in the QIAS 2025 and QIAS 2026 shared tasks \cite{bouchekif-etal-2025-qias, bouchekif2026qias2026overviewshared} employed RAG-based approaches to retrieve information from trusted Islamic sources, such as IslamWeb, for answering general Islamic questions and Islamic inheritance cases. Similarly, \cite{alan2025improvingllmreliabilityrag, bhatia2026ragagenticragfaithful, fanar-sadiq, eltanbouly2026groundingarabicllmsdoha} integrated RAG and agentic RAG pipelines to support grounded and faithful Islamic question answering. RAG has also been applied to specialized Islamic domains, including Seerah question answering using Hadith sources \cite{seerahgpt}, Quran-related tasks such as the IslamicEval 2025 shared task \cite{mubarak-etal-2025-islamiceval}, Quranic studies \cite{Khalila2025}, and Tafsir retrieval \cite{al-azani-etal-2025-ontologyrag}. Beyond RAG-based applications, recent benchmarks have also evaluated LLM knowledge and reasoning across Islamic domains, including IslamicMMLU \cite{abdelaal2026islamicmmlubenchmarkevaluatingllms}, MAWARITH \cite{bouchekif2026mawarithdatasetbenchmarklegal}, IslamicTurathBench \cite{gaben2026islamicturathbenchmultitaskmultidisciplinebenchmark}, and IslamicLegalBench \cite{elmahjub2026islamiclegalbenchevaluatingllmsknowledge}.

Despite this growing body of work in Islamic AI, standalone retrieval evaluation remains comparatively limited. Existing work has centered on the Qur'an and, more recently, Hadith. The Qur'an QA shared tasks established verse-level passage retrieval as a benchmark task \cite{malhas-etal-2022-quran, malhas-etal-2023-quran}, and a substantial body of work has since improved Qur'anic retrieval through domain-adapted language models, data augmentation, and fine-tuning \cite{pavlova-2023-leveraging, basem-2024-quran, pavlova-2025-multi}, as well as multilingual and cross-lingual retrieval over Qur'anic translations \cite{pavlova-makhlouf-2025-efficient, Comparative_quran}. Retrieval over related religious sources has also been explored, including Tafsir \cite{JADS1000} and, in the IslamicEval 2025 shared task, joint Qur'an and Hadith passage retrieval \cite{mubarak-etal-2025-islamiceval, basheer-etal-2025-burhan, al-adel-etal-2025-burhanai, amin-etal-2025-nur, elkoumy-etal-2025-tce}. Across this body of work, retrieval evaluation in the Islamic domain remains concentrated on the Qur'an, Hadith, and Tafsir. Retrieval-level evaluation for fiqh texts, particularly in the context of answer-bearing and madhhab-aware retrieval, remains underexplored.

\section{Retrieval Pipeline} \label{pipeline}

The retrieval pipeline consists of two stages: an optional madhhab-filtering stage followed by a retrieval stage. For a given Arabic fiqh question, the pipeline returns a ranked list of candidate passages from the corpus.

\subsection{Madhhab Filtering} \label{madhhab}
Many fiqh questions are specific to a particular madhhab, and the ruling adopted by one school may differ from that of another. Because passages from different schools often discuss the same issue using highly similar language, unconstrained retrieval may return school-inconsistent rulings alongside answer-bearing passages from the target school. To address this issue, we introduce a madhhab-filtering step before retrieval.

School detection is performed using a rule-based lexicon containing madhhab names, adjectival and plural forms, and the names of founding scholars. For example, \textarabic{الحنفية}, \textarabic{أحناف}, and \textarabic{أبي حنيفة} are all mapped to the \textarabic{حنفي} school. Questions without any school reference are treated as general fiqh questions.  An evaluation of the detector on the test collection is provided in Appendix~\ref{app:madhhab-detection}.

When a school is detected, retrieval is restricted to passages associated with that madhhab. If multiple schools are mentioned, passages associated with any of them are considered. Passages that are not associated with a specific madhhab are excluded when filtering is applied. The filtering stage reduces cross-school confusion before ranking, while the retrieval stage ranks only the candidate passages allowed by the filter.

\subsection{Retrieval}
After applying the optional madhhab filter, passages are retrieved using one of three retrieval strategies. \textbf{Dense retrieval:} The query is encoded using a dense embedder and matched against passage embeddings using cosine similarity. \textbf{Lexical retrieval:} The query is matched against passages using BM25 over a sparse lexical representation. \textbf{Hybrid retrieval:} Dense and lexical retrieval are performed independently, and their rankings are combined using Reciprocal Rank Fusion (RRF) \cite{RRF_paper}.

\section{Experimental setup} \label{ES}
This section describes the resources and setup, including the corpus, fine-tuning data, test collection, models, and evaluation settings.

\subsection{Retrieval Corpus}
Our retrieval corpus consists of approximately 356K chunks extracted from 100 classical fiqh books. The chunks were constructed systematically by segmenting each \textit{mas'alah} (an individual juristic issue or question) based on the structural organization of the original books. This approach preserves the contextual boundaries of each fiqh issue while maintaining meaningful retrieval units. The resulting chunks have an average length of 255 words. Each chunk is assigned one or more madhhab labels based on the author's madhhab and references to madhahib within the text. Additional details on corpus construction and madhhab annotation are provided in Appendix \ref{A:madhhab-annotation}.

\subsection{Fine-tuning Data} \label{ft-data}
Existing retrieval models are not adapted to the specialized vocabulary and concepts of fiqh. To adapt retrieval models to this domain, we constructed a dataset of training triplets. Each triplet consists of a question, an answer-bearing passage, and a hard negative passage.

Questions were generated from a subset of five fiqh books using Gemini 2.5 Flash and paired with the source chunk from which they were generated to form question-positive pairs, and at most one question was retained per chunk. Negative passages were selected using the corpus metadata. For each question, a negative chunk was sampled from the same section as the positive passage but from a different subsection, producing topically related yet non-answer-bearing negatives.

The final dataset contains 19,319 question-positive-negative triplets covering multiple madhahib and topics. Additional details on data generation and selection are provided in Appendix \ref{A:ft-dataset-details}.

\subsection{Test Collection} \label{TC}

Our test collection consists of 503 human-authored questions written with reference to specific juristic scenarios discussed in the fiqh books. Each scenario is defined by a set of conditions under which a ruling is sought. Of these, 123 are \textit{general} questions that are not specific to any school, while 380 are \textit{madhhab-based} questions distributed across the four Sunni schools. The questions span a range of fiqh topics, including acts of worship (\textarabic{عبادات}), transactions (\textarabic{معاملات}), religious etiquette (\textarabic{آداب شرعية}), and customs (\textarabic{عادات}).

\textbf{Relevance criterion.} We adopt a binary answer-bearing notion of relevance: a chunk is relevant only if it explicitly provides the ruling requested by the question. Chunks that are merely on-topic, discuss only part of the issue, or omit the requested ruling are considered non-relevant. For madhhab-based questions, only answer-bearing chunks belonging to the requested school are considered relevant.

\textbf{Relevance labeling.} Initial relevance judgments were derived from the juristic scenario associated with each question and subsequently verified through LLM-assisted review and human annotation. Disputed cases were reviewed by an annotator holding a master's degree in Islamic Studies, with final checking and corrections by an Islamic Studies expert holding a PhD in the field. Human judgments were treated as definitive.  The training and evaluation data were constructed from disjoint sets of chunks to prevent train--test contamination. Additional details are provided in Appendix~\ref{appendix:test-collection}.

\subsection{Models}
We evaluate eight off-the-shelf embedding models, grouped into three categories based on their training background. The \textbf{multilingual} category includes BGE-M3 \cite{bge-m3}, Multilingual-E5-large-instruct \cite{e5-large}, Arctic-Embed-L~v2.0 \cite{arctic-2}, Nomic-Embed-text-v2 \cite{nomic-ai}, and Qwen3-Embedding-0.6B \cite{qwen3embedding}. The \textbf{general Arabic} category is represented by Arabic-Triplet-Matryoshka-V2 (ATM2) \cite{ATM2}. Finally, the \textbf{domain-adapted Arabic} category consists of two legal-domain variants, Muffakir~V1 and Muffakir~V2 \cite{muffakir}, derived from different backbones.

For fiqh-specific fine-tuning, we select one representative model from each category: BGE-M3, ATM2, and Muffakir~V1. Full architectural details are provided in Appendix~\ref{app:models}.

\subsection{Metrics}

We evaluate retrieval at $k{=}5$ using \textbf{MRR@5}, \textbf{nDCG@5}, and \textbf{Hit@5} as primary metrics. We additionally report \textbf{MAP@5} and \textbf{Recall@5} and more values of $k$ in Appendix \ref{cutoffs_results}. Statistical significance is assessed using two-sided Wilcoxon signed-rank tests on per-question retrieval scores with Holm--Bonferroni correction to account for multiple comparisons. Additional details are provided in Appendix~\ref{app:eval-details}.

\subsection{Implementation Details}
Fine-tuning is performed using Sentence-Transformers~\cite{Sentence-transformer} with Matryoshka Loss~\cite{MRL} and Multiple Negatives Ranking Loss (MNRL), using the fiqh triplets described in Section~\ref{ft-data}. For hybrid retrieval, the top 50 candidates from both the dense retriever and BM25 are fused using Reciprocal Rank Fusion (RRF) with a rank constant of 60. Additional training and implementation details are provided in Appendix~\ref{A:imp_details}.
\section{Evaluation Results} \label{results}
Our evaluation is organized around two research questions: retriever selection (RQ1) and madhhab-aware retrieval (RQ2).

\subsection{RQ1: Retriever selection} \label{rq1}
RQ1 investigates three aspects of retriever selection for answer-bearing fiqh retrieval. Specifically, we examine the effectiveness of existing embedding models (RQ1.1), the impact of fiqh-specific fine-tuning (RQ1.2), and the relative performance of dense, lexical, and hybrid retrieval strategies (RQ1.3). Full results are reported in Table~\ref{tab:model_incremental}.

\begin{table}[t]
\centering
\small
\setlength{\tabcolsep}{3pt}
\begin{tabular}{lccc}
\toprule
\textbf{Model} & \textbf{NDCG@5} & \textbf{MRR@5} & \textbf{Hit@5} \\
\midrule
\textbf{Muffakir V1} & & & \\
\quad Base  & 0.3585 & 0.5238 & 0.6720 \\
\quad + FT & 0.3764 & 0.5534 & 0.7058 \\
\qquad └ + BM25 & \textbf{0.4007} & \textbf{0.5739} & \textbf{0.7237} \\
\midrule
\textbf{ATM2} & & & \\
\quad Base  & 0.2959 & 0.4407 & 0.5805 \\
\quad + FT & 0.3638 & 0.5377 & 0.6899 \\
\qquad └ + BM25 & 0.3886 & 0.5642 & 0.7217 \\
\midrule
\textbf{BGE-M3} & & & \\
\quad Base  & 0.2965 & 0.4425 & 0.5746 \\
\quad + FT & 0.3187 & 0.5014 & 0.6402 \\
\qquad └ +BM25 & 0.3696 & 0.5375 & 0.6859 \\
\midrule
\textbf{Nomic} & & & \\
\quad Base  & 0.2974 & 0.4435 & 0.6044 \\
\quad + BM25  & 0.3597 & 0.5307 & 0.6859 \\
\midrule
\textbf{mE5-large} & & & \\
\quad Base  & 0.3160 & 0.4822 & 0.6044 \\
\quad + BM25  & 0.3612 & 0.5246 & 0.6839 \\
\midrule
\textbf{Arctic} & & & \\
\quad Base  & 0.3097 & 0.4642 & 0.6024 \\
\quad + BM25  & 0.3702 & 0.5128 & 0.6720 \\
\midrule
\textbf{Muffakir V2} & & & \\
\quad Base  & 0.2274 & 0.3514 & 0.4791 \\
\quad + BM25  & 0.3380 & 0.4979 & 0.6382 \\
\midrule
\textbf{Qwen3-0.6B} & & & \\
\quad Base  & 0.2605 & 0.3963 & 0.5408 \\
\quad + BM25  & 0.3332 & 0.4755 & 0.6262 \\
\midrule
\textbf{BM25} & 0.2897 & 0.4140 & 0.5746 \\
\bottomrule
\end{tabular}
\caption{Evaluation results across the retrieval pipeline. Each tier shows the effect of fiqh fine-tuning (FT) and hybrid retrieval (+BM25).}
\label{tab:model_incremental}
\end{table}

\textbf{\textit{RQ1.1:}} Among retrievers, Muffakir achieved the strongest performance across all metrics, reaching an MRR@5 of 0.5238. Compared to its base model, ATM2 (MRR@5 = 0.4407), this improvement suggests that adaptation to legal-domain data transfers effectively to fiqh, likely because both domains share formal language, specialized terminology, and structured reasoning patterns.
The next strongest group consists of the multilingual models E5, Arctic, Nomic, and BGE-M3. Despite not being specifically designed for Arabic Islamic text, they achieved competitive results, with E5 and Arctic reaching MRR@5 scores of 0.4822 and 0.4642, respectively. This suggests that large-scale multilingual pretraining provides strong cross-domain and cross-lingual transfer capabilities that extend to Classical Arabic retrieval.

Qwen3, the only decoder-based embedding model evaluated, achieved an MRR@5 of 0.3963, performing below the BM25 baseline (0.4140). Muffakir V2 performed even worse (MRR@5 = 0.3514), despite being trained on additional adaptation data. Since Muffakir V1 and V2 are derived from different backbones and adaptation pipelines, this result suggests that retrieval performance depends not only on adaptation but also on the underlying architecture and adaptation strategy.

Overall, the results highlight the importance of both model architecture and adaptation data for answer-bearing fiqh retrieval, while showing that multilingual models remain competitive despite the absence of explicit adaptation to Classical Arabic or Islamic text.

\textbf{\textit{RQ1.2}:} To evaluate the impact of domain-specific adaptation, we fine-tuned three models representing different model categories: Muffakir, ATM2, and BGE-M3. Since Muffakir is derived from ATM2 through an intermediate legal-domain adaptation stage, comparing their fiqh-fine-tuned versions allows us to assess whether prior adaptation to a related domain remains beneficial after direct fiqh adaptation.

All three models improved after fine-tuning on the fiqh dataset. ATM2 exhibited the largest gains, with MRR@5 increasing from 0.4407 to 0.5377 (+0.097, +22\% relative), followed by BGE-M3, which improved from 0.4425 to 0.5014 (+0.059, +13.3\%). Muffakir achieved the highest overall dense retrieval performance after fine-tuning (MRR@5 = 0.5534), improving from 0.5238 (+0.030, +5.7\%). Although its relative gain was smaller, Muffakir already started from the strongest baseline, leaving less room for improvement. These results demonstrate that both Arabic-specific and multilingual retrievers can effectively adapt to specialized fiqh retrieval tasks.

A particularly interesting pattern emerges when examining the adaptation path from ATM2 to Muffakir. Legal-domain adaptation increased MRR@5 from 0.4407 to 0.5238 (+0.083), while subsequent fiqh fine-tuning further increased performance to 0.5534. However, after both ATM2 and Muffakir were fine-tuned on fiqh data, the performance gap between them shrank to only 0.0157 MRR@5, and the two AraBERT-based models became statistically indistinguishable across all metrics. This suggests that while adaptation to a related domain provides a useful starting point, direct in-domain adaptation captures most of the task-specific knowledge required for fiqh retrieval.

The results also highlight the importance of the adaptation data itself. While fiqh fine-tuning substantially improved BGE-M3, Muffakir V2, another BGE-M3-derived model adapted primarily on Arabic multidomain data, achieved an MRR@5 of only 0.3514, performing below the original base model. Although further investigation is needed to explain this behavior, the contrast suggests that adaptation to Arabic alone is insufficient; the adaptation data must align with the retrieval task. In-domain answer-bearing fiqh supervision appears considerably more valuable than broad language-specific adaptation.

\textbf{\textit{RQ1.3}:} Although dense retrieval consistently outperformed the BM25 baseline, lexical matching remained complementary even to the strongest dense retrievers. BM25 retrieved a relevant chunk for 54 questions (10.7\%) missed by fine-tuned ATM2 and 48 questions (9.5\%) missed by fine-tuned Muffakir, motivating the evaluation of a hybrid configuration.

The effectiveness of hybrid retrieval depended strongly on the quality of the dense retriever. The largest gains were observed for weaker models: Muffakir V2 improved from 0.3514 to 0.4979 MRR@5 (+0.1465), while other mid- and lower-performing retrievers also showed substantial improvements. This suggests that lexical and semantic signals are genuinely complementary when the dense representation is less effective.

For the strongest fine-tuned retrievers, however, the gains were small and not statistically significant. Fine-tuned ATM2 improved from 0.5377 to 0.5642 MRR@5 (+0.0265), while fine-tuned Muffakir improved from 0.5534 to 0.5739 (+0.0205). Among the fine-tuned models, only BGE-fiqh showed a significant hybrid improvement, and only on nDCG@5 among the main metrics.

Importantly, the limited gains do not indicate that the hybrid failed to exploit BM25's complementary signal. Of the questions missed by dense retrieval but retrieved by BM25, the hybrid recovered 70.4\% for ATM2 and 77.1\% for Muffakir. However, these improvements were partially offset by regressions on questions already answered by the dense model, resulting in only modest net gains. In effect, hybrid retrieval mainly redistributes hits near the rank-5 boundary rather than substantially improving ranking quality.

\begin{table}[t]
\centering
\small
\setlength{\tabcolsep}{1pt}
\begin{tabular}{lcccccc}
\toprule
\multirow{2}{*}{\textbf{Model}} & \multicolumn{3}{c}{\textbf{MRR@5}} & \multicolumn{3}{c}{\textbf{Latency (ms)}} \\
\cmidrule(lr){2-4}\cmidrule(lr){5-7}
 & Dense & Hybrid & $\Delta$ & Dense & Hybrid & $\times$ \\

\midrule
BGE-fiqh        & 0.501 & 0.538 & $+$0.036 & 119 & 306 & 2.6 \\
ATM2-fiqh       & 0.538 & 0.564 & $+$0.026 & 109 & 294 & 2.7 \\
Muffakir-fiqh   & 0.553 & 0.574 & $+$0.021& 121 & 279 & 2.3 \\
\bottomrule
\end{tabular}
\caption{Effectiveness and efficiency of dense-only vs.\ hybrid retrieval. For the three fiqh fine-tuned models, hybrid fusion roughly doubles latency for a small, non-significant MRR@5 gain.}
\label{tab:hybrid_cost}
\end{table}

The practical value of hybrid retrieval therefore depends on the strength of the underlying dense retriever. As shown in Table~\ref{tab:hybrid_cost}, hybrid retrieval increased latency by approximately 2.3--2.7$\times$ while providing only small improvements for the strongest models. We therefore conclude that hybrid retrieval is highly beneficial for weaker retrievers, but its advantage becomes limited once strong fiqh-specific dense retrievers are used. Balancing effectiveness and efficiency, fine-tuned ATM2 and Muffakir operating in dense-only mode represent the most practical choice.

\subsection{RQ2: Madhhab-aware retrieval }
RQ2 investigates two aspects of madhhab-aware retrieval: differences in retrieval performance between general and madhhab-based questions (RQ2.1) and the effect of restricting retrieval to the relevant madhhab (RQ2.2).

\textbf{\textit{RQ2.1}:} Retrieval performance differs substantially between the two question types (Table \ref{tab:gen_vs_madhhab}). Fine-tuned ATM2 achieves an MRR@5 of 0.706 on general questions compared to 0.483 on madhhab-based questions, and fine-tuned Muffakir achieves 0.695 and 0.508, respectively, which is a gap of roughly 0.20 MRR@5 in both cases. Similar trends are observed across all evaluated models.

A major reason for this gap is the difference in the number of answer-bearing chunks available for each question type. General questions have, on average, more than four times as many relevant chunks as madhhab-based questions (41.7 vs. 9.7). When many ground truth chunks exist, the probability that at least one appears in the top retrieved results is higher, leading to higher Hit@5 and MRR@5 scores. This is supported by Recall@5, which is higher for madhhab questions (0.165) than for general questions (0.075), indicating that the retriever retrieves a larger fraction of the smaller relevant set despite its lower MRR@5.

However, the gap cannot be explained solely by the size of the relevant set. Fiqh fine-tuning consistently improves retrieval on madhhab-based questions while having a much smaller effect on general questions. For example, ATM2 improves from 0.372 to 0.483 MRR@5 on madhhab-based questions, reducing the performance gap between general and madhhab-based questions from 0.284 to 0.223. Similar reductions are observed for Muffakir and BGE-M3. Fine-tuning therefore substantially improves performance on madhhab-based questions and narrows the gap with general questions. Nevertheless, a considerable difference remains because the limited number of answer-bearing chunks for madhhab-based questions imposes a constraint that model adaptation alone cannot fully overcome.
\begin{table}[t]
\centering
\small
\setlength{\tabcolsep}{2pt}
\begin{tabular}{lcccc}
\toprule
\multirow{2}{*}{\textbf{Model}} & \multicolumn{2}{c}{\textbf{MRR@5}} & \multicolumn{2}{c}{\textbf{Recall@5}} \\
\cmidrule(lr){2-3}\cmidrule(lr){4-5}
 & General & Madhhab & General & Madhhab \\
\midrule
ATM2-fiqh      & 0.706 & 0.483 & 0.075 & 0.165 \\
Muffakir-fiqh  & 0.695 & 0.508 & 0.084 & 0.178 \\
BGE-fiqh      & 0.670 & 0.447 & 0.068 & 0.141 \\
\bottomrule
\end{tabular}
\caption{Retrieval performance on general and madhhab-based questions. General questions achieve higher MRR@5, while madhhab-based questions achieve higher Recall@5.}
\label{tab:gen_vs_madhhab}
\end{table}
\textbf{\textit{RQ2.2:}}
We introduce a madhhab-filtering stage that restricts retrieval to the passages of the relevant school whenever a question is identified as school-specific. This stage produces one of the largest performance improvements observed in our pipeline.

\begin{figure}
    \centering
    \includegraphics[width=1\linewidth]{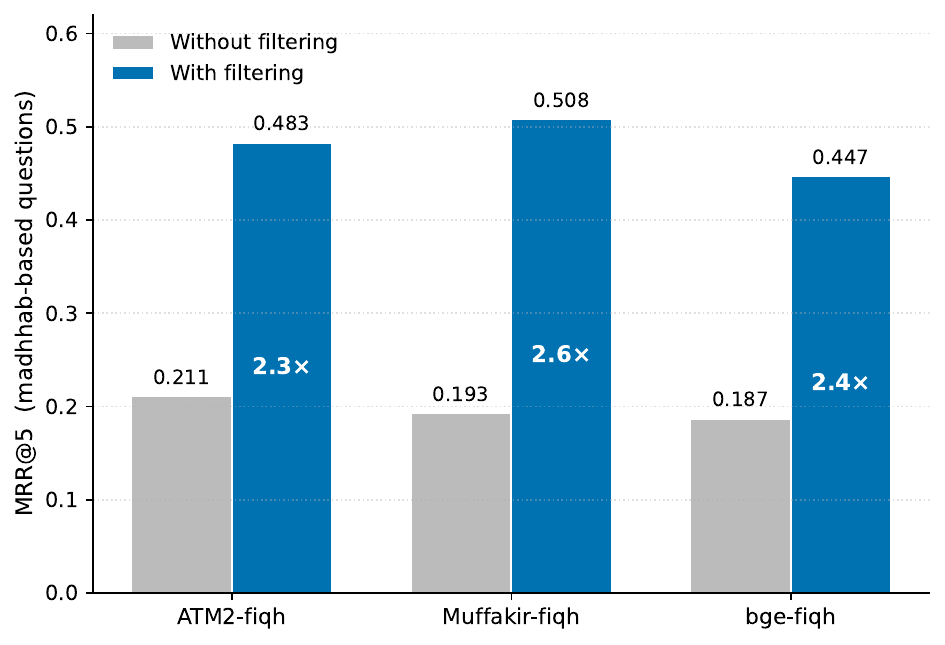}
    \caption{Effect of madhhab filtering on madhhab questions for the three fine-tuned models. }
    \label{fig:madhhab-impact}
\end{figure}

On madhhab-based questions, filtering more than doubles retrieval effectiveness (Figure~\ref{fig:madhhab-impact}). MRR@5 rises from 0.211 to 0.483 for fine-tuned ATM2, from 0.193 to 0.508 for fine-tuned Muffakir, and from 0.187 to 0.447 for fine-tuned bge, with comparable gains in Hit@5 (all $p<10^{-25}$, Wilcoxon signed-rank). This improvement exceeds that of fiqh fine-tuning or hybrid fusion, yet comes at negligible cost, as the routing is rule-based and adds no learned components.
This gain results from reducing the cross-school confusion described in Section~\ref{madhhab}. When filtering is removed, Recall@5 on madhhab-based questions drops sharply (e.g., for fine-tuned ATM2, from 0.165 to 0.061), as passages from other schools compete with and displace the answer-bearing chunks from the top retrieved results.

\begin{table*}[!ht]
\centering
\small
\begin{tabularx}{\textwidth}{
@{}
p{0.05\textwidth}
p{0.22\textwidth}
p{0.30\textwidth}
X
@{}
}\toprule
\textbf{Ctg.} & \textbf{Question} & \textbf{Retrieved Passage} & \textbf{Why it is Incorrect} \\
\midrule

CM &
\textarabic{كيف يكون بر الوالدين بعد موتهما؟ عند الشافعية} &
\textarabic{وأفضل البر بر الوالدين بالإحسان إليهما وفعل ما يسرهما من الطاعة لله تعالى وغيرها...} &
Retrieves a general ruling on honoring parents while missing the condition \textit{after their death}.
\\
\midrule

QTM &
\textarabic{متى يضمن المالك في المزارعة} &
\textarabic{ المزارعة : الشركة في الزرع,ويقال : الشركة في الحرث. وعقدها غير لازم قبل البذر ونحوه...
} &
Provides contract definition and validity information rather than the requested liability conditions.
\\

\midrule

CI &
\textarabic{ما قول الحنفية في: حكم قتل الأصل للفرع المحارب  من الكفار؟} &
\textarabic{...لا يجوز له قتل أصله الحربي إلا دفعا عن نفسه , وإن خاف رجوعه ضيق عليه وألجأه ليقتله غيره...} &
Matches the general issue but misses the specific roles and circumstances
\\
\midrule

TM &
\textarabic{لو اقترض أحد الزوجين من الآخر، فهل يحق له الحبس بهذا القرض؟} &
\textarabic{في المقارض يأخذ من رجل آخر مالا قراضا قلت : أرأيت إن أخذ رجل مالا قراضا من رجل... } &
Confuses similar terms: \textarabic{القراض} (profit-sharing) vs. \textarabic{القرض}(loan). \\

\bottomrule
\end{tabularx}
\caption{Representative retrieval errors. CM: Condition Mismatch; QTM: Question-Type Mismatch; CI: Context Misinterpretation; TM: Topic Mismatch.}
\label{tab:error-analysis-examples}
\end{table*}

\section{Error Analysis} \label{EA}
To better understand retrieval failures, we analyze them from both quantitative and qualitative perspectives. We focus on questions that received a Hit@5 score of zero for the both top-performing models, namely fine-tuned ATM2 and fine-tuned Muffakir. This allows us to examine failure cases that persist even under the strongest retrieval configurations.
\subsection{Quantitative Analysis}

\paragraph{Recall vs.\ ranking failures.}
We decompose each missed question ($Hit@5=0$) into two mechanisms: either no answer-bearing passage is retrieved into the candidate pool (a \emph{recall} failure), or an answer-bearing passage is retrieved but ranked outside the top five results (a \emph{ranking} failure). Across both retrievers, recall failures are more common: for roughly two-thirds of misses, no answer-bearing passage appears in the top 20 candidates (62\% for fine-tuned ATM2 and 64\% for fine-tuned Muffakir). The remaining cases are ranking failures. This indicates that retrieval failures are driven primarily by difficulty retrieving answer-bearing passages into the candidate pool rather than ranking them.

\paragraph{Separability.}
Similarity scores provide further insight into the ranking failures. On missed questions, the top-ranked passage still receives a high similarity score (0.618 for fine-tuned ATM2 and 0.640 for fine-tuned Muffakir), only slightly below successful queries (0.685 and 0.702, respectively). Retrieval failures therefore do not arise from the absence of semantically related passages, but from difficulty distinguishing answer-bearing passages from highly similar non-answer-bearing ones.

This is reflected in the similarity margin between the highest-scoring answer-bearing passage and the highest-scoring non-answer-bearing passage. On missed questions, the margin is negative ($-0.079$ for ATM2 and $-0.081$ for Muffakir), indicating that a non-answer-bearing passage receives a higher score than any available answer-bearing passage. Even on successful queries, the margin remains small (+0.019 and +0.023), and among cases where an answer-bearing passage reaches the candidate pool, a non-answer-bearing passage still receives a higher score in half of the queries (52\% and 48\%). These results suggest that answer-bearing and topically related non-answer-bearing passages remain only weakly separable in the embedding space.

\subsection{Qualitative Analysis}

The quantitative analysis shows that retrieval failures are often caused by confusion among semantically related passages. To understand the nature of these confusions, we manually examined questions that received a Hit@5 score of zero for both top-performing models and analyzed the passages commonly retrieved by both to focus the analysis on shared retrieval failures rather than model-specific errors.

We identified four categories of retrieval errors that are largely driven by the characteristics of fiqh texts. Since a single failure may exhibit multiple patterns and the boundaries between categories are not always clear-cut, we present these categories qualitatively rather than as a quantitative distribution. Representative examples are shown in Table~\ref{tab:error-analysis-examples}.

\textbf{Condition Mismatch}: Fiqh rulings are often highly dependent on specific conditions and circumstances. Retrieval failures occur when the model overlooks a condition mentioned in the question and retrieves a more general ruling, retrieves a ruling restricted to conditions not present in the question, or retrieves a passage discussing a different condition within the same underlying issue.

\textbf{Question-Type Mismatch}: Fiqh questions may ask about different aspects of a topic, such as the ruling, underlying reasoning, validity conditions, obligations, or exceptions. In some cases, the retrieved passage discusses the correct topic and scenario but answers a different type of question, making it irrelevant despite its high semantic similarity.

\textbf{Context Misinterpretation}: Some fiqh cases involve multiple actors, actions, or relationships whose roles determine the ruling. The retriever may identify the correct topic and conditions but fail to distinguish the context, retrieving passages that describe a similar scenario with reversed or altered roles.

\textbf{Topic Mismatch}: In these cases, the retriever fails to capture the intended meaning of the question and retrieves passages from a different fiqh topic altogether. This may occur because certain terms appear similar despite referring to distinct jurisprudential concepts.

\paragraph{Observations.}
Taken together, the quantitative and qualitative analyses show that retrieval failures in fiqh are mainly caused by the difficulty of distinguishing between closely related passages rather than by retrieving completely unrelated content. The quantitative results show that missed questions often still receive high similarity scores, but the retrieved passages are difficult to separate from the relevant ones because answer-bearing passages are only weakly distinguished from topically similar non-answer-bearing passages. The qualitative analysis further explains the source of this confusion: most errors involve passages that discuss the same underlying issue but differ in a critical aspect, such as a condition, the type of information requested, or the roles of the involved actors. Even topic mismatch cases often result from confusion between closely related jurisprudential terms rather than a complete failure to identify the general area of the question. These findings indicate that current retrieval models capture broad semantic similarity effectively but struggle with the fine-grained distinctions that determine the applicability of fiqh rulings. Therefore, improving fiqh retrieval requires retrieval strategies that better capture conditions, contextual roles, and question types, rather than relying on improved ranking, as many failures occur because the relevant passage is not retrieved into the candidate pool in the first place.

\section{Conclusion}
\label{conclusion}

This paper presented a retrieval-focused evaluation of answer-bearing fiqh retrieval. Our results show that fiqh fine-tuning improves dense retrievers, while hybrid retrieval adds useful lexical signals but gives only modest gains for the strongest fine-tuned models. Madhhab-aware filtering provides the largest improvement on school-specific questions by reducing cross-school confusion.

Our error analysis shows that remaining failures are mainly caused by difficulty distinguishing answer-bearing passages from topically similar passages. These findings suggest that future work should focus on harder fiqh-specific negatives and retrieval models that better capture fine-grained juristic distinctions.

\section*{Limitations}

Our study has several limitations. First, our relevance judgments are binary: a passage is treated as either answer-bearing or non-relevant. This does not capture partial relevance, such as passages that provide part of a ruling or state a ruling without its supporting conditions. A graded relevance scheme would represent such cases more faithfully but would require substantially more annotation effort.

Second, the relevance annotation followed an expert review and checking procedure rather than independent double annotation. While this provided expert verification of disputed cases, it does not allow us to report inter-annotator agreement. Future work could strengthen the validation process by incorporating multiple LLM judgments and independent annotations from multiple Islamic Studies experts.

Third, the fine-tuning questions were automatically generated using an LLM. Although the generation prompts were developed with Islamic Studies experts and sampled outputs were manually inspected, the full synthetic dataset was not exhaustively human-reviewed. The generated data may therefore contain occasional wording artifacts or imperfect question--passage pairs.

Fourth, our madhhab detection is lexicon-based and therefore susceptible to surface-level collisions. For example, the word \textarabic{المالك} (the owner) contains the substring \textarabic{مالك} (M\=alik), causing one general question in our test collection to be incorrectly flagged as Maliki-specific. Such cases were rare and did not affect the reported results, but a morphology-aware detector would reduce these errors.

Finally, our evaluation is restricted to the retrieval stage and to models of at most approximately 0.6B parameters. We do not measure how retrieval improvements translate to downstream answer quality, nor do we evaluate larger retrievers or alternative architectures such as late-interaction models. In addition, the retrieval errors identified in our analysis suggest that future work should explore harder fiqh-specific negatives and retrieval representations that better capture fiqh semantics, particularly distinctions involving conditions, contextual roles, and closely related jurisprudential concepts.

\section*{Acknowledgment}
This study was conducted as part of a project funded by the ARG grant (ARG01-0524-230318), awarded by the Qatar Research, Development, and Innovation Council (QRDI).

\bibliography{bibliography}
\appendix
\section{Appendix}

\subsection{Data Construction Details}
\label{appendix-A}

\subsubsection{Madhhab Annotation} \label{A:madhhab-annotation}

Each chunk is assigned one or more madhhab labels. For books associated with a specific madhhab, all extracted chunks inherit that madhhab as their primary label. In addition, chunks may receive additional labels when they explicitly discuss the opinions of other madhahib. For example, a Hanafi text comparing the Hanafi and Shafi'i positions would be assigned both Hanafi and Shafi'i labels, with Hanafi retained as the primary label. This allows comparative fiqh discussions to be retrieved when searching for either school.

Madhhab mentions are detected through explicit references to madhhab names as well as references to recognized scholars associated with a particular school. For books that are not attributed to a specific madhhab, labels are assigned solely based on the madhhab mentions identified within the text.

Table \ref{tab:madhhab-distribution} reports the distribution of chunks across madhahib.

\begin{table}[t]
\centering

\begin{tabular}{lr}
\toprule
\textbf{Madhhab} & \textbf{\# Chunks} \\
\midrule
Hanafi &  100,950\\
Maliki &  61,302\\
Shafi'i &  95,780\\
Hanbali &  46,409\\
Other Schools &  17,345\\
General / Comparative Fiqh &  34,331\\
\midrule
\textbf{Total} & \textbf{356,117} \\
\bottomrule
\end{tabular}
\caption{Distribution of chunks by primary madhhab label in the retrieval corpus.}
\label{tab:madhhab-distribution}
\end{table}

\subsubsection{Fine-tuning data details}
\label{A:ft-dataset-details}

Questions and answers were generated from a subset of five fiqh books using Gemini 2.5 Flash. The model was provided with fiqh chunks containing the original in-text structural annotations and was instructed to generate question-answer pairs grounded in the supplied passage. The generation prompts were developed in consultation with experts in Islamic studies and refined through multiple rounds of review. Several generated batches were manually inspected during development to verify that the questions were faithful to the source passages, reflected the intended ruling, and were suitable for retrieval training. 

To reduce redundancy, at most one question was retained from each chunk. Hard negatives were selected from the same section as the positive passage but from a different subsection, producing topically related yet non-answer-bearing negatives. This strategy helps the retriever to distinguish between passages discussing closely related fiqh issues rather than relying on broad topical differences. As a sample-based validation of this procedure, 100 negative passages were manually reviewed, and all were confirmed to be non-answer bearing for their corresponding questions. This check was performed on a sample and does not constitute manual validation of the full negative dataset.

The final dataset contains 19,319 triplets distributed across the Sunni schools (5,040 Shafi'i, 5,001 Hanafi, 4,879 Hanbali, and 4,399 Maliki). An example triplet is shown in Table~\ref{triplet-example}.

\begin{table*}[t]
\centering
\small
\begin{tabular}{p{0.15\textwidth} p{0.78\textwidth}}
\toprule
\textbf{Element} & \textbf{Content} \\
\midrule

Question &
\textarabic{
ما حكم الصلح عن شيء مغصوب تالف بأكثر من قيمته (كألفين مقابل ألف) أو بعرض (سلعة)؟ }\\
\midrule

Pos. Passage &\textarabic{
"غصب ثوباً أو عبداً قيمته ألف فاستهلكه، فصالحه على ألفين أو عرض جاز ... وكذا الصلح بعرض صح وإن كان قيمته أكثر من قيمة المغصوب التالف لعدم الربا."}\\
\midrule

Neg. Passage &\textarabic{
"وركنه الإيجاب والقبول ... ويصح الصلح مع الإقرار، والإقرار أقوى من الشهادة ..." }\\
\midrule

Why Negative &
The passage discusses the general validity and requirements of \textit{\d{s}ul\d{h}} (settlement), making it topically related to the question. However, it does not address settlement over a destroyed usurped item, nor does it provide the specific ruling requested by the question. \\
\bottomrule

\end{tabular}
\caption{Example fine-tuning triplet. The negative passage is topically related to the question but does not provide the requested ruling.}
\label{triplet-example}
\end{table*}

\subsubsection{Test Collection Construction}
\label{appendix:test-collection}

\textbf{Initial Relevance Pool}: Questions were authored with reference to specific juristic scenarios discussed in the source fiqh books. Each scenario is defined by a particular set of conditions and circumstances under which a ruling is sought. For each question, all chunks associated with the corresponding scenario were initially included in the relevance pool, as these chunks collectively discuss the case from which the question was derived.

\textbf{Answer-Bearing Verification}: Although chunks associated with the same scenario generally discuss the same juristic case, not all of them necessarily provide the specific ruling requested by a question. To identify such cases, all chunks in the initial relevance pool were assessed according to the answer-bearing relevance criterion described in Section~\ref{TC}.

To identify potentially missed relevant passages, we additionally retrieved the top 100 BM25 passages for each question. These passages were combined with the original scenario pool and screened by Gemini~2.5~Flash according to the answer-bearing relevance criterion.

LLM judgments were used only to identify candidate additions and removals. Whenever the LLM judgment disagreed with the original scenario-based label, the chunk was reviewed by a human annotator. Chunks were removed from the relevance set only when the annotator confirmed that they did not provide the requested ruling. Similarly, retrieved passages outside the original scenario pool were added only when the annotator confirmed that they contained the requested ruling. Human judgments were treated as definitive in all cases.

To prevent train--test contamination, we ensured that no chunk included in the test collection appeared in the fine-tuning dataset. The training and evaluation data are therefore disjoint at the chunk level.

Table~\ref{tab:test-collection-examples} shows representative examples of a general question and a madhhab-based question together with excerpts from answer-bearing passages judged relevant under our annotation criteria.

\begin{table*}[t]
\centering
\small
\begin{tabular}{p{0.10\linewidth} p{0.30\linewidth} p{0.50\linewidth}}
\toprule
Type & Question & Relevant Passage (Excerpt) \\
\midrule

General &
\textarabic{ما حكم الالتفات في الصلاة؟}
&
\textarabic{لا خلاف بين الفقهاء في كراهة الالتفات في الصلاة ... والكراهة مقيدة بعدم الحاجة أو العذر.}
\\

\midrule

Madhhab-based &
\textarabic{وفق المذهب المالكي، ماذا يحدث لعقد الاستصناع إذا مات أحد المتعاقدين؟}
&
\textarabic{الصناع أحق بما أسلم إليهم في الموت والفلس ما كان بأيديهم ...}
\\

\bottomrule
\end{tabular}
\caption{Examples of questions in the test collection and corresponding answer-bearing passages. Only excerpts containing the relevant ruling are shown.}
\label{tab:test-collection-examples}
\end{table*}

\textbf{Annotation Statistics}: A total of 3,657 disputed chunk labels were reviewed by human annotators. Human annotators agreed with the Gemini judgment in 80.0\% of reviewed disagreements and with the original label in the remaining 20.0\%.

Among the reviewed disagreements, annotators confirmed the removal of 1,612 chunks from the original scenario pools because they did not directly answer the corresponding question, and added 1,313 answer-bearing chunks retrieved outside the original scenario.

\subsection{Models Details}
\label{app:models}

Table~\ref{tab:models} summarizes the architectural characteristics of the eight evaluated embedding models. Most are encoder-based retrievers, while Qwen3-Embedding-0.6B is the only decoder-based model. ATM2 serves as the base model for Muffakir~V1, whereas Muffakir~V2 is derived from BGE-M3, allowing comparison of different legal-domain adaptation paths. The models also vary in parameter count, embedding dimensionality, context length, and pooling strategy.

\begin{table*}
\centering
\small
\setlength{\tabcolsep}{3pt}
\begin{tabular}{llllllll}
\toprule
\textbf{Model} &
\textbf{Category} &
\textbf{Arch.} &
\textbf{Backbone} &
\textbf{Params} &
\textbf{Dim} &
\textbf{Ctx} &
\textbf{Pooling} \\
\midrule

BGE-M3
& Multi.
& Enc
& XLM-R-large
& 568M
& 1024
& 8192
& CLS \\

Qwen3-Emb-0.6B
& Multi.
& Dec
& Qwen3-0.6B
& 600M
& 1024
& 32768
& Last tok \\

Arctic-Embed-L-v2
& Multi.
& Enc
& XLM-R-large
& 568M
& 1024
& 8192
& CLS \\

Nomic-Embed-v2
& Multi.
& Enc (MoE)
& Nomic BERT
& 475M*
& 768
& 512
& Mean \\

mE5-large-inst
& Multi.
& Enc
& XLM-R-large
& 560M
& 1024
& 512
& Mean \\

ATM2
& Arabic
& Enc
& AraBERTv02
& 135M
& 768
& 512
& Mean \\

Muffakir V1
& Domain Ar.
& Enc
& AraBERTv02
& 135M
& 768
& 512
& Mean \\

Muffakir V2
& Domain Ar.
& Enc
& XLM-R-large
& 362M
& 1024
& 8192
& CLS \\

\bottomrule
\end{tabular}
\caption{Summary of the eight retrieval models evaluated. Multi. = multilingual, Enc = encoder, Dec = decoder, and Ctx = maximum context length. *Nomic-Embed-v2 is a mixture-of-experts model with 475M total parameters (305M active).}
\label{tab:models}
\end{table*}

\subsection{Metrics and Statistical Testing}
\label{app:eval-details}

\subsubsection{Evaluation Metrics}

Retrieval effectiveness is evaluated at rank $k=5$, corresponding to the number of passages typically supplied to a generator in a RAG pipeline.
\textbf{MRR@5} (Mean Reciprocal Rank): Measures how highly the first answer-bearing passage is ranked.
\textbf{nDCG@5} (Normalized Discounted Cumulative Gain): Measures the overall ranking quality of relevant passages, giving higher weight to passages appearing near the top of the ranking.
\textbf{Hit@5}: Measures whether at least one answer-bearing passage appears among the top five retrieved results.
\textbf{MAP@5} (Mean Average Precision): Measures ranking quality by averaging precision over all relevant passages retrieved within the top five results.
\textbf{Recall@5}: Measures the fraction of all answer-bearing passages retrieved within the top five results. Because many questions have more relevant passages than the retrieval cutoff, Recall@5 is used primarily as a diagnostic measure rather than a primary evaluation metric.

For latency, we report the mean per-query wall-clock retrieval time in milliseconds. Timing wraps only the retrieval call and excludes data loading and index construction, thereby measuring the end-to-end cost of the retrieval pipeline itself.

\subsubsection{Statistical Testing}

All statistical comparisons use a two-sided Wilcoxon signed-rank test applied to per-question retrieval scores. This test was selected because retrieval metrics are bounded, non-normal, and computed on paired observations (the same questions evaluated under different retrieval configurations).

Holm--Bonferroni correction was applied within each analysis to account for multiple comparisons. Unless otherwise stated, significance levels are reported as: $^{*}$ $p < 0.05$, $^{**}$ $p < 0.01$, $^{***}$ $p < 0.001$, ns: not significant.

Four groups of statistical comparisons were performed: (1) \textbf{Fine-tuning Effect}: Comparison of each base model with its fiqh-fine-tuned counterpart.(2) \textbf{Fine-Tuned Model Comparison}: Pairwise comparisons among the fiqh-fine-tuned models.(3) \textbf{Hybrid Retrieval}: Comparison of dense-only and hybrid (dense + BM25) retrieval configurations.(4) \textbf{Madhhab Filtering}: Comparison of retrieval performance with and without madhhab-based filtering. Detailed results for all statistical tests are reported in Appendix~\ref{app:significance}.

\subsubsection{Implementation Details} \label{A:imp_details}
\textbf{BM25}: BM25 retrieval is implemented using the Qdrant/FastEmbed BM25 model. Before indexing, Arabic diacritics were removed from all corpus texts, as queries are typically written without diacritics; this preprocessing reduces term mismatches between queries and indexed passages.
\textbf{Fine-tuning: }Models are trained for 3 epochs with a batch size of 128, a learning rate of $2\times10^{-5}$, and a fixed seed of 42. We reserve 5\% of the training triplets for validation and select the checkpoint with the highest validation performance.
\textbf{Retrieval:} all passage embeddings are stored in a Qdrant collection using named vectors, allowing the same corpus to be searched with different embedding models. All experiments were conducted on a single NVIDIA RTX 5880 Ada GPU.

\subsection{Statistical Significance}
\label{app:significance}

We assess statistical significance using the two-sided Wilcoxon signed-rank test over per-question scores. The test is paired (the same questions are scored under both conditions) and non-parametric (per-  question retrieval metrics are bounded and non-normal). Within each analysis we apply Holm--Bonferroni correction across all reported (model~$\times$~metric) comparisons. We denote $p<0.05$ ($^{*}$), $p<0.01$ ($^{**}$), $p<0.001$ ($^{***}$), and ``ns'' otherwise.

\begin{table}[h]
\centering\small
\begin{tabular}{lccc}
\toprule
Model & $\Delta$MRR@5 & $\Delta$nDCG@5 & $\Delta$Hit@5 \\
\midrule
ATM2-fiqh     & $+0.097^{***}$ & $+0.068^{***}$ & $+0.109^{***}$ \\
BGE-fiqh      & $+0.059^{**}$  & $+0.022$\,ns   & $+0.066^{**}$  \\
Muffakir-fiqh & $+0.030$\,ns   & $+0.018$\,ns   & $+0.034$\,ns   \\
\bottomrule
\end{tabular}
\caption{Significance of fiqh fine-tuning (fine-tuned $-$ base, dense, $n=503$).}
\label{tab:sig-ft}
\end{table}

\paragraph{Effect of fiqh fine-tuning.}
Table~\ref{tab:sig-ft} compares each base model with its fiqh-fine-tuned
version. Fine-tuning is statistically significant for ATM2 across all metrics and for BGE-M3 on MRR@5 and Hit@5. For Muffakir~V1, which starts from the strongest baseline, the gains are positive but not significant, consistent with its limited headroom.

\paragraph{Pairwise comparison of fine-tuned models.}
Table~\ref{tab:sig-pair} reports pairwise tests among the three fine-tuned models. The two AraBERT-based models, ATM2-fiqh and Muffakir-fiqh, are not significantly different on any metric, supporting the convergence reported in Section~\ref{rq1}. Both differ from BGE-fiqh on nDCG@5.

\begin{table}[h]
\centering\small
\setlength{\tabcolsep}{1pt}

\begin{tabular}{lccc}
\toprule
Comparison & MRR@5 & nDCG@5 & Hit@5 \\
\midrule
ATM2-fiqh vs Muffakir-fiqh & ns & ns & ns \\
ATM2-fiqh vs BGE-fiqh      & ns & $^{***}$ & ns \\
Muffakir-fiqh vs BGE-fiqh  & $^{*}$ & $^{***}$ & $^{*}$ \\
\bottomrule
\end{tabular}
\caption{Pairwise significance among fine-tuned models (dense), Holm-corrected.}
\label{tab:sig-pair}
\end{table}

\paragraph{Dense vs.\ hybrid retrieval.}
Table~\ref{tab:sig-hybrid} reports hybrid-vs-dense differences for the
fine-tuned models. For ATM2-fiqh and Muffakir-fiqh, no metric shows a significant hybrid gain. For BGE-fiqh, the gain is significant on nDCG@5, MAP@5, and Recall@5, but not on MRR@5 or Hit@5. This supports the conclusion that hybrid fusion offers little benefit to strong fine-tuned retrievers.

\begin{table}[h]
\centering\small
\setlength{\tabcolsep}{1pt}

\begin{tabular}{lccccc}
\toprule
Model & MRR@5 & nDCG@5 & Hit@5 & MAP@5 & Rec@5 \\
\midrule
ATM2-fiqh     & ns & ns & ns & ns & ns \\
Muffakir-fiqh & ns & ns & ns & ns & ns \\
BGE-fiqh      & ns & $^{***}$ & ns & $^{***}$ & $^{***}$ \\
\bottomrule
\end{tabular}
\caption{Significance of hybrid vs.\ dense retrieval (fine-tuned models, $n=503$), Holm-corrected.}
\label{tab:sig-hybrid}
\end{table}

\paragraph{Madhhab filtering.}
Table~\ref{tab:sig-filter} reports the effect of madhhab filtering on the 380 madhhab-based questions. Filtering yields highly significant improvements for all three fine-tuned models across every metric ($p<10^{-24}$ throughout), confirming that restricting retrieval to the relevant school is the single largest source of improvement in our pipeline.

\begin{table}[h]
\centering\small
\begin{tabular}{lccc}
\toprule
Model & $\Delta$MRR@5 & $\Delta$nDCG@5 & $\Delta$Hit@5 \\
\midrule
ATM2-fiqh     & $+0.272^{***}$ & $+0.202^{***}$ & $+0.311^{***}$ \\
Muffakir-fiqh & $+0.315^{***}$ & $+0.219^{***}$ & $+0.297^{***}$ \\
BGE-fiqh      & $+0.260^{***}$ & $+0.171^{***}$ & $+0.292^{***}$ \\
\bottomrule
\end{tabular}
\caption{Significance of madhhab filtering (filtered $-$ unfiltered) on the 380 madhhab-based questions. All metrics significant at $p<10^{-24}$.}
\label{tab:sig-filter}
\end{table}

\begin{table}[h]
\centering\small
\begin{tabular}{llccc}
\toprule
Model & $k$ & MRR@$k$ & nDCG@$k$ & Hit@$k$ \\
\midrule
\multirow{5}{*}{ATM2-fiqh} & 1 & 0.4414 & 0.4414 & 0.4414 \\
 & 3 & 0.5215 & 0.3911 & 0.6183 \\
 & 5 & 0.5377 & 0.3638 & 0.6899 \\
 & 10 & 0.5468 & 0.3317 & 0.7535 \\
 & 20 & 0.5513 & 0.3189 & 0.8171 \\
\midrule
\multirow{5}{*}{Muffakir-fiqh} & 1 & 0.4573 & 0.4573 & 0.4573 \\
 & 3 & 0.5401 & 0.3979 & 0.6481 \\
 & 5 & 0.5534 & 0.3764 & 0.7058 \\
 & 10 & 0.5623 & 0.3454 & 0.7694 \\
 & 20 & 0.5661 & 0.3328 & 0.8231 \\
\midrule
\multirow{5}{*}{BGE-fiqh} & 1 & 0.4215 & 0.4215 & 0.4215 \\
 & 3 & 0.4854 & 0.3480 & 0.5686 \\
 & 5 & 0.5014 & 0.3187 & 0.6402 \\
 & 10 & 0.5103 & 0.2892 & 0.7078 \\
 & 20 & 0.5142 & 0.2777 & 0.7654 \\
\bottomrule
\end{tabular}
\caption{Retrieval effectiveness at different $k$ for the fine-tuned models. MRR@$k$ and Hit@$k$ increase with larger candidate pools, nDCG@$k$ decreases because it is normalized against the full set of relevant passages per question.}
\label{tab:k-sweep}
\end{table}

\subsection{Additional Results} \label{add_results}
\subsubsection{Performance at Different Retrieval Cutoffs} \label{cutoffs_results}

Table~\ref{tab:k-sweep} reports retrieval effectiveness for the three best-performing fine-tuned retrievers at different retrieval cutoffs. While the main paper focuses on $k{=}5$, which matches the number of passages typically supplied to a generator in a RAG pipeline, these results illustrate how retrieval effectiveness changes as additional passages are retrieved.

\subsubsection{Madhhab Detection Evaluation}
\label{app:madhhab-detection}

We evaluated the rule-based madhhab detector on the test collection to verify that the filtering stage correctly identifies school-specific questions. On the 380 madhhab-based questions, the detector achieved perfect recall (1.000), precision of 0.985, and F1 of 0.992. The correct madhhab was detected for every school-specific question, with exact single-school matches in 375 out of 380 cases (98.7\%). On the 123 general questions, only one false school detection occurred.

Most errors were false positives where an additional madhhab was detected alongside the expected one. Per-school recall was 1.000 for all four madhahib, while precision was 1.000 for Hanbali, 0.990 for Hanafi, 0.990 for Shafi'i, and 0.958 for Maliki. These results indicate that the filtering stage is reliable for routing madhhab-based questions, while the remaining errors are rare and mainly due to surface-form ambiguity.

\end{document}